\documentclass[letter,superscriptaddress,twocolumn, pra,showkeys,showpacs]{revtex4}

	\usepackage{amsmath}
	\usepackage{makeidx}
	\usepackage{amsfonts}
	\usepackage[ansinew]{inputenc}
	\usepackage[usenames,dvipsnames]{pstricks}
    \usepackage{graphicx}
    \usepackage{float}
    \usepackage{subfigure}
	\usepackage{pst-grad} 
	\usepackage{pst-plot} 
	\usepackage{sidecap}
	\usepackage[makeroom]{cancel}
	\usepackage{textcomp}

	\usepackage[colorlinks,hyperindex]{hyperref}
	
	\hypersetup
	{
		colorlinks,%
		citecolor=black,%
		linkcolor=black,%
		urlcolor=black,%
	}

	\newcommand{\ket}[1]{\left| #1 \right\rangle}
	\newcommand{\bra}[1]{\left\langle #1 \right|}
	
	\newcommand{\braket}[2]{\left\langle #1 \right. \left| #2 \right\rangle}

\makeindex

\begin{document}

\title{Dynamics and protection of entanglement in $n$-qubit systems within Markovian and non-Markovian environments }

\author{Alireza Nourmandipour}
\email{anoormandip@stu.yazd.ac.ir}
\affiliation{Atomic and Molecular Group, Faculty of Physics, Yazd University,  89195-741 Yazd, Iran}
\author{Mohammad Kazem Tavassoly}
\email{mktavassoly@yazd.ac.ir}
\affiliation{Atomic and Molecular Group, Faculty of Physics, Yazd University,  89195-741 Yazd, Iran}
\affiliation{Photonic Research Group, Engineering Research Center, Yazd University,  89195-741 Yazd, Iran}
\author{Morteza Rafiee}
\email{m.rafiee178@gmail.com}
\affiliation{Department of Physics, Shahrood University of Technology, 3619995161 Shahrood, Iran}

\date{today}

\begin{abstract}
We provide an analytical investigation of the pairwise entanglement dynamics for a system, consisting an arbitrary number of qubits dissipating into a common and non-Markovian environment for both weak and strong coupling regimes. In  the latter case, a revival of pairwise entanglement due to the memory depth of the environment is observed. The leakage of photons into a continuum state is assumed to be the source of dissipation. We show that for an initially Werner state, the environment washes out the pairwise entanglement, but a series of non-selective measurements can protect the relevant entanglement. On the other hand, by limiting the number of qubits initially in the superposition of single excitation, a stationary entanglement can be created between qubits initially in the excited and ground states. Finally, we determine the stationary distribution of the entanglement versus the total number of qubits in the system.
\end{abstract}

\pacs{03.65.Ud, 03.67.Mn, 03.65.Yz}
\keywords{$n$-qubit systems; Dissipative systems; Quantum entanglement; Quantum Zeno effect}

\maketitle

\section{Introduction}
Entanglement is one of the marvellous aspects of quantum mechanics which has no corresponding classical equivalence \cite{Horodecki2009QuantumEntanglement}. This notion has been widely used due to its important role as a resource for quantum information processing  \cite{Ekert1991Cryptography,Braunstein1995,Mattle1996,Abdi2012,Ou2012,Muarao1999}. At the early stages of quantum information studies, the decoherence induced by the environment was recognized as the main obstacle in preserving the entanglement. Therefore, it seemed quite logical  to avoid  interaction with environment as much as possible. Altogether, the possibility of achieving long-time entangled states has been put forward in numerous works that focused on the generation of entangled states via coupling qubits to a common and dissipative environment \cite{Plenio2002,Mancini2015,Angelakis2009,Krauter2011}. Quite remarkably, it has been also shown that, the environment can play a constructive role in establishing entanglement between subsystems even without any interaction among them \cite{Nourmandipour2015,Memarzadeh2013,Maniscalco2008,Memarzadeh2011,Rafiee2012,Vasile2009,Genkin2011,Gao2010,Paz2008,Horhammer2008}. Actually, the common environment provides an indirect interaction between the subsystems which leads to establish entanglement among them. The possibility of environment-induced entanglement for systems composed of only two subsystems has been considered by many authors \cite{Nourmandipour2015,Maniscalco2008}. On the other hand, protecting of entanglement in real devices is crucial for practical quantum information processing purposes. Therefore, many attempts have been devoted to fight against the deterioration of entanglement  under impact of environment \cite{Kim2012,Ghanbari2014}. In this regard, it is shown that, the evolution of an unstable quantum system can be slowed down or even freezed if the mentioned system is observed continuously. This method is known as the quantum Zeno effect (QZE) \cite{Kofman2000}. Actually, it is not necessary that the state of the system remains frozen in a single state, but it could just evolve in a multidimensional subspace, namely the Zeno subspace \cite{Facchi2002}.

In this paper, we consider a model in which an arbitrary number of qubits interact with a cavity field and the cavity mode itself interacts with a set of continuum harmonic oscillators. We then intend to study the possibility of environment induced entanglement generation between these qubits with the environment outside the Markovian regime for both weak and strong couplings corresponding to the bad and good cavity limits, respectively. We obtain the exact dynamics of various pairwise entanglements as a function of the environment correlation time for both coupling regimes. In particular, in the strong coupling, we show that how the entanglement revivals and oscillations can be induced due to the long memory of the reservoir. We then show that, when all of the qubits are initially in a superposition of the single excited states with the same probability (i.e., a Werner state), the pairwise entanglement decreases as time goes on for any value of system size, $n$, and no stationary entanglement can be achieved in neither of coupling regimes. However, as will be seen, a series of nonselective measurements can quench the decay of entanglement. On the contrary, when the initial state is considered as a superposition of one excitation of two arbitrary qubits, the environment not only establishes entanglement between various pairs of qubits, but also makes it persists up to stationary state. The achievable stationary entanglement is independent of the environmental variables and only depends on the system size and  initial conditions. The entanglement generated for pairs of qubits initially in the ground states is negligibly smaller than pairs of initially excited and non-excited qubits.

The rest of the paper is organized as follows: In Sec. \ref{sec.model} we introduce the relevant Hamiltonian describing our system. In Sec. \ref{sec:werner} by considering the initial state of qubits as a Werner state, we examine the effect of environment on the pairwise entanglement. Furthermore, we preserve entanglement from environment decoherence by quantum Zeno effect. In Sec. \ref{sec.twoqubit}, it is illustrated that  limiting the number of qubits in the superposition of one excitation could lead to stationary entanglement. Finally, the paper ends with  concluding remarks in Sec. \ref{sec.conclusion}.

\section{Model}\label{sec.model}
The system under consideration consists of $n$ qubits with associated Hilbert space ${\cal H}\simeq{\cal C}^{2\otimes n}$ dissipating into a common environment. Let us $\left\lbrace \ket{0},\ket{1}\right\rbrace ^{\otimes n}$ be the orthonormal basis in which, $\ket{0}$ ($\ket{1}$) is the ground (excited) single qubit state. We model our dissipative system as a high-Q cavity in which the qubits interact with the single-mode cavity field which is characterized by annihilation operator $\hat{a}$ and frequency $\omega_c$ via coupling constant $g$ and the cavity field itself interacts with an external field which is considered as a set of continuum harmonic oscillators with annihilation and creation operators $\hat{B}(\eta)$ and $\hat{B}^{\dagger}(\eta)$ at mode $\eta$ through coupling coefficient $G(\eta)$. From this point of view, one can find that photons in the cavity can leak out to a continuum of states, which is the source of dissipation. We show that this model leads to a Lorentzian spectral density which implies the nonperfect reflectivity of the cavity mirrors. The correlation between the $i$th qubit and the cavity field is characterized by the terms like $g\left( \hat{\sigma}_+^{(i)}\hat{a} + \hat{\sigma}_-^{(i)}\hat{a}^{\dagger}\right) $ in which $ \hat{\sigma}_{+}^{(i)}$ ($\hat{\sigma}_{-}^{(i)}$) is the raising (lowering) operator of the $i$th qubit, and the interaction between the cavity and the external fields can be governed by the Hamiltonian
\begin{equation}
 \begin{aligned}
 \hat{H}_{\text{I}}&= \omega_c \hat{a}^{\dagger} \hat{a}  \ + \ \int_0^{\infty}\! \eta \hat{B}^{\dagger}(\eta)\hat{B}(\eta) \, \mathrm{d}\eta \ \\ &+ \ \int_0^{\infty}\! \left\lbrace  G(\eta)\hat{a}^{\dagger}\hat{B}(\eta)\ + \ \text{H.c.} \right\rbrace \, \mathrm{d}\eta.
  \end{aligned}
  \label{eq:int}
  \end{equation}
In \cite{Nourmandipour2015} we demonstrated that, by assuming that the surrounding environment possesses such a narrow bandwidth  that only a particular mode of the cavity can be excited \cite{Dutra2005}, one is able to extend integrals over $\eta$ back to $-\infty$ and take $G(\eta)$ as a constant (equal to $\sqrt{\kappa/\pi}$). This procedure allows us to diagonalize the Hamiltonian (\ref{eq:int}) using the dressed operator $\hat{A}(\omega)=\alpha(\omega)\hat{a}+\int\! \beta(\omega,\eta)\hat{B}(\eta) \, \mathrm{d}\eta$, where $\alpha(\omega)$ and $\beta(\omega,\eta)$ (in general $\in \mathbb{C}$) are obtained such that $\hat{A}(\omega)$ is an annihilation operator which obeys the commutation relation with its conjugate as $\left[ \hat{A}(\omega),\hat{A}^{\dagger}(\omega^{'})\right]=\delta(\omega-\omega^{'})$ \cite{Fano1961,Barnett2002}. The bosonic operator $\hat{a}$ can be shown to be a linear combination of the dressed operator $\hat{A}(\omega)$ as follow \cite{Dutra2005,Nourmandipour2015}:
\begin{equation}
\hat{a}=\int\! \alpha^{*}(\omega)\hat{A}(\omega) \, \mathrm{d}\omega,
\label{eq:operatorA}
\end{equation}
with
\begin{equation}
\alpha(\omega)=\frac{\sqrt{\kappa/\pi}}{\omega-\omega_c+i\kappa}. \label{eq:alpha}
\end{equation}
From this point of view, one can dedicate that, the qubits dissipate into a common environment which is now described by the annihilation and creation operators $\hat{A}(\omega)$ and $\hat{A}^{\dagger}(\omega)$, respectively.
Thus, the interaction between a generic qubit with the surrounding environment is governed by terms like $g\int\!\left(\hat{\sigma}_+^{(i)}\alpha^{*}(\omega)\hat{A}(\omega)+\text{H.c.}\right)\mathrm{d}\omega$.  Henceforth, the total Hamiltonian of our system can be rewritten in terms of the dressed operators as follows
\begin{equation}
\hat{H}=\hat{H}_\text{S}+ \hat{H}_{\text{Env}} + \hat{H}_{\text{Int}},
\label{eq:finalhamiltonian}
  \end{equation}
where $\hat{H}_{\text{S}}$ is the Hamiltonian of the qubits coupled, via the interaction Hamiltonian $\hat{H}_{\text{Int}}$ to the common environment with the qualifier Hamiltonian $\hat{H}_{\text{Env}}$. In the dipole and  rotating-wave approximations, and in units of $\hbar=1$, they can be written as
\begin{subequations}
\label{eq:H2}
\begin{eqnarray}
\hat{H}_\text{S}&=&\dfrac{\omega_{\text{qb}}}{2}\sum_{i=1}^{n}\hat{\sigma}_{z}^{(i)}, \label{eq:S} \\
\hat{H}_{\text{Env}}&=&\int\! \omega \hat{A}^{\dagger}(\omega)\hat{A}(\omega) \, \mathrm{d}\omega, \label{eq:Env} \\
\hat{H}_{\text{Int}}&=& g\sum_{i=1}^{n}\int\!\left(\hat{\sigma}_+^{(i)}\alpha^{*}(\omega)\hat{A}(\omega)+\text{H.c.}\right) \, \mathrm{d}\omega,  \label{eq:intt}
\end{eqnarray}
\end{subequations}
in which, $\hat{\sigma}_{z}^{(i)}$ is inversion population operators of the $i$th qubit. In writing Hamiltonian (\ref{eq:H2})  we have assumed  that all qubits have the same resonance frequency, namely, $\omega_{\text{qb}}$ and also the coupling constant between all qubits and the cavity field be the same, say $g$.

The time-dependent Schrödinger equation with Hamiltonian (\ref{eq:finalhamiltonian}) can be solved  when  the environment initially is  in a vacuum state and the system of qubits are in an arbitrary superposition of single excitation of qubits. We specially shall address the cases in which the system of qubits are in a Werner state and in a superposition of one excitation of two arbitrary qubits  in the  two next sections, separately. It should be noted that, solving the time-dependent Schrödinger equation analytically with arbitrary initial state seems to be a very hard task, if not impossible.

\section{System of $n$-Qubits Initially in a Werner State}\label{sec:werner}
In this section, we assume that the set of qubits initially be in a Werner state \cite{Werner1989} and there is no excitation in the cavity before the occurrence of interaction. Therefore
\begin{equation}
\ket{\psi(0)}=\ket{w}\ket{\boldsymbol{0}}_{R},
\label{eq:initialwerner}
\end{equation}
in which $\ket{w}:=1/\sqrt{n}\sum_{k=0}^{n}\ket{1_k}$ is the Werner state where $\ket{1_k}\equiv\ket{0_1,\cdots,1_k,\cdots,0_n}$ and $\ket{\boldsymbol{0}}_{R}=\hat{A}(\omega)\ket{1_{\omega^{'}}}\delta(\omega-\omega^{'})$ is the multi-mode vacuum state, where $\ket{1_{\omega}}$ is the multi-mode state representing one photon at frequency $\omega$ and vacuum state in all other modes. Consequently, the time evolution of the state of the system may be proposed as:
\begin{equation}
\ket{\psi(t)}={\cal E}(t)e^{i\omega_{\text{qb}}t}\ket{w}\ket{\boldsymbol{0}}_{R}
+\int\! \Lambda_{\omega}(t)e^{i\omega t}\ket{1_{\omega}}\ket{G} \, \mathrm{d}\omega,
\label{eq:statewerner}
\end{equation}
in which $\ket{G}:=\ket{0}^{\otimes n}$ and
\begin{equation}
\left| {\cal E}(t)\right| ^2\equiv\text{P}_0(t)=\left| \braket{\psi(0)}{\psi(t)}\right| ^2
\label{eq:suramp}
\end{equation}
is the survival probability of the initial state. Following the approach which recently has been presented in \cite{Nourmandipour2015} and after lengthy but straightforward manipulations, the following  integro-differential equation for the amplitude ${\cal E}(t)$ can be obtained:
\begin{equation}
\label{eq:diffu}
\dot{{\cal E}}(t)=-\int_{0}^{t}\! f(t-t_1){\cal E}(t_1) \, \mathrm{d}t_1,
\end{equation}
in which the correlation function $f(t-t_1)$ reads as
\begin{equation}
f(t-t_1)=\int\! \, \mathrm{d}\omega J(\omega) e^{i\delta_{\omega}(t-t_1)}  , \label{eq:f}
\end{equation}
where $\delta_{\omega}=\omega_{\text{qb}}-\omega$. Here, in deriving (\ref{eq:diffu}), we have assumed that the qubits interact with the cavity field in the exact resonance condition, i.e. $\omega_{\text{qb}}-\omega_c=0$. At this stage, it should be noted that according to Eq. (\ref{eq:alpha}), the spectral density obeys the Lorentzian distribution, i.e.,
\begin{equation}
\label{eq:spcden}
J(\omega)=\dfrac{1}{\pi}\dfrac{ng^2\kappa}{ (\omega-\omega_c)^2+\kappa^2}.
\end{equation}
This result has been indeed directly obtained from our modelling of dissipative cavity and implies the nonperfect reflectivity of the cavity mirrors \cite{Breuer2002}. This leads to an exponentially decaying correlation function, with $\kappa$ as the decay rate factor of the cavity; consequently, the cavity correlation time is  $\tau_B\approx\kappa^{-1}$. On the other hand, it can be shown that the relaxation time $\tau_R$ over which the state of the system consisting of only one qubit changes is $\tau_R\approx g^{-1}$ \cite{Bellomo2007}.  Altogether, by choosing special values of $\kappa$, it is possible to extract the ideal cavity and the Markovian limits. The former is obtained when $\kappa\rightarrow 0$, which leads to $J(\omega)=ng^2\delta(\omega-\omega_c)$ corresponding to a constant correlation function. In this situation, the system reduces to a $n$-qubit Jaynes-Cummings model \cite{Tavis1968} with the vacuum Rabi frequency $\Omega_R=\sqrt{n}g$. Moreover, for small correlation times and by taking $\kappa$ much larger than any other frequency scale, the Markovian regime may be obtained. For the other generic values of $\kappa$, the model interpolates between these two limits.

The Laplace transform technique helps to solve the integro-differential equation (\ref{eq:diffu}) for the surviving amplitude which arrives us at:
 \begin{equation}
{\cal E}(t)=e^{-\kappa t/2}\left( \cosh{\left(\Omega_n t/ 2 \right)} +\dfrac{\kappa}{\Omega_n}\sinh{\left( \Omega_n t/ 2\right) }\right),  \label{eq:survivalamplitude}
 \end{equation}
where $\Omega_n=\sqrt{\kappa^2-4g^2 n}$.
As is seen, the obtained solution is quite exact with no approximation. From Eq. (\ref{eq:survivalamplitude}), it is clear that letting $t$ to tend to infinity, ${\cal E}(\infty)\longrightarrow 0$. Therefore, looking at (\ref{eq:statewerner}), one can realise that $\ket{\psi(\infty)}\propto \ket{G}$, which implies that when all qubits are initially in a superposition of single excited states with the same probability, no stationery entanglement can be achieved.
\subsection{Dynamics of Entanglement}
\begin{figure}[h!]
\centering
\subfigure[\label{Fig1a} \ Bad cavity limit, $R=0.1$.]{\includegraphics[width=0.35\textwidth]{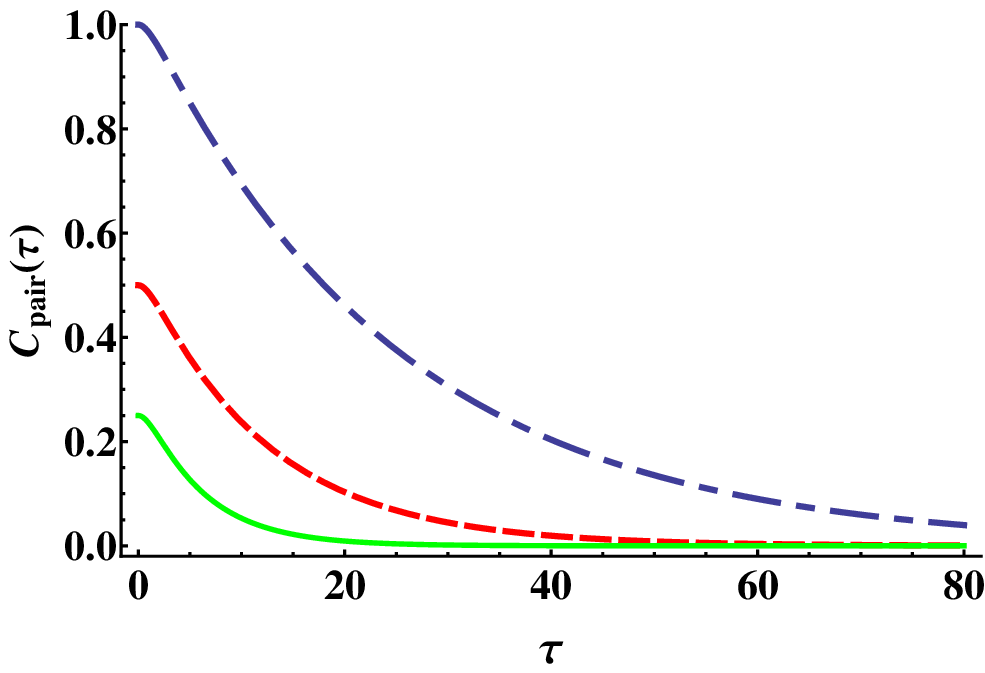}}
\hspace{0.05\textwidth}
\subfigure[\label{Fig1b} \ Good cavity limit, $R=10$.]{\includegraphics[width=0.35\textwidth]{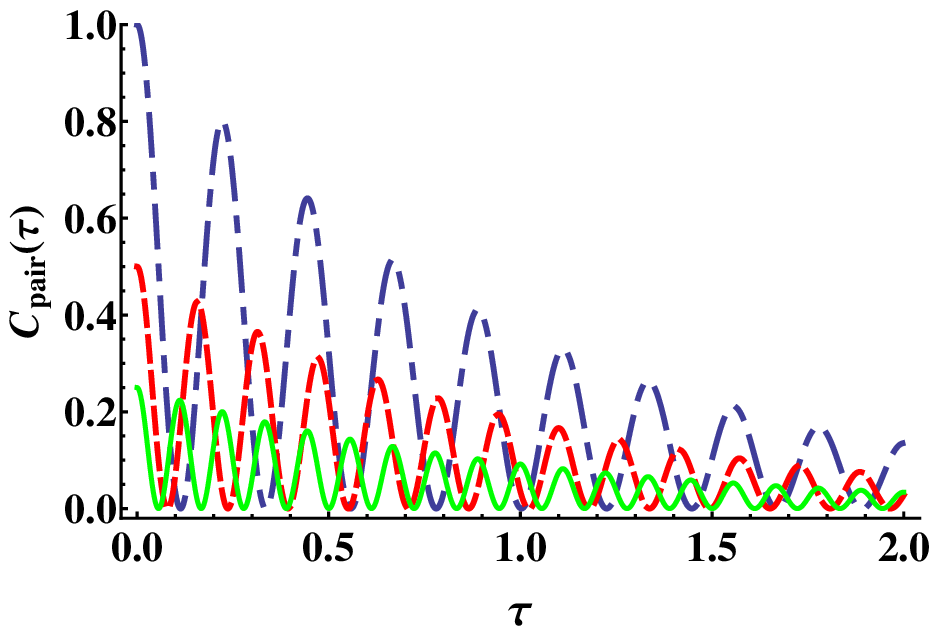}}

\caption{(Color online) Pairwise concurrence ${\cal C}_{\text{pair}}$ as function of $\tau$ when the initial state of the system is a Werner state, in the bad cavity limit, i.e. $R=0.1$ (top plot) and good cavity limit, $R=10$ (bottom plot) with $n=2$ (dot-dashed blue line), $n=4$ (dashed red line), and $n=8$ (solid green line).} \label{Fig1}
   \end{figure}

Using (\ref{eq:statewerner}) the explicit form of the reduced density operator for the system of qubits can be derived by tracing over environment variables as follows:
\begin{equation}
\rho(t)=\left| {\cal E}(t)\right| ^2\ket{w}\bra{w}+\left( 1-\left| {\cal E}(t)\right| ^2\right) \ket{G}\bra{G}.
\label{eq:redwer}
\end{equation}
The coefficient of the last term is arisen from the fact that $\text{Tr}(\rho(t))=1$.
In what follows, we use concurrence as a suitable measure to quantify the amount of entanglement between various pairs of qubits which is defined as \cite{Wootters1998}
\begin{equation}
{\cal C}(t)=\max\left\lbrace 0, \sqrt{\ell_1}- \sqrt{\ell_2}- \sqrt{\ell_3}- \sqrt{\ell_4}\right\rbrace,
\label{eq:con}
\end{equation}
where $\left\lbrace \ell_j\right\rbrace _{j=1}^4$ are the eigenvalues (in decreasing order) of the Hermitian matrix
$\rho\left(\sigma_1^y\otimes\sigma_2^y\rho^{*}\sigma_1^y\otimes\sigma_2^y\right)$ with $\rho^*$ as the complex conjugate of $\rho$ in the standard basis and $\sigma_k^y:=i(\sigma_k-\sigma_k^\dag)$ in the same basis. The concurrence varies between 0 (when the qubits are separable) and 1 (when they are maximally entangled). To analyse the pairwise entanglement between any two generic qubits, we compute partial trace of (\ref{eq:redwer}) over all other qubits and obtain the following reduced density operator:
\begin{equation}
\rho_{\text{pair}}(\tau)=\begin{pmatrix}
 0 & 0 & 0 & 0 \\
 0 & \dfrac{\left| {\cal E}(\tau)\right| ^2}{n} & \dfrac{\left| {\cal E}(\tau)\right| ^2}{n} & 0 \\
 0 & \dfrac{\left| {\cal E}(\tau)\right| ^2}{n} & \dfrac{\left| {\cal E}(\tau)\right| ^2}{n} & 0 \\
 0 & 0 & 0 & 1-\dfrac{2\left| {\cal E}(\tau)\right| ^2}{n}
 \end{pmatrix},
 \label{eq:rwkl}
\end{equation}
where the relevant concurrence reads as ${\cal C}_{\text{pair}}(\tau)=2\left| {\cal E}(\tau)\right| ^2/n$, in which the dimensionless scaled time $\tau$ has been defined as $\tau=\kappa t$.  Keeping in mind (\ref{eq:suramp}), it is readily found that ${\cal C}_{\text{pair}}(\tau)=2\text{P}_0(\tau)/n$, which implies that the pairwise concurrence directly depends on the survival probability of the initial state. Looking at Eq. (\ref{eq:survivalamplitude}), it is clear that the weak and strong coupling regime can be distinguished. The weak (strong) coupling regime can be obtained by $R^2<(4n)^{-1}$ ($R^2>(4n)^{-1}$), in which we have defined the dimensionless parameter  $R=g/\kappa$. The quantity ${\cal C}_{\text{pair}}(\tau)$ is shown in Fig. \ref{Fig1} in both regimes. In weak coupling regime, the relaxation time is greater than the reservoir correlation time and the behaviour of $\text{P}_0$ is essentially a Markovian exponential decay. The concurrence disappears faster when the system size is larger. The strong coupling regime is more rich and represents the revival and oscillation of entanglement. This revival phenomenon is due to the long memory of the reservoir.  In this case, the reservoir correlation time is greater than the relaxation time and non-Markovian effects become dominant. The concurrence periodically vanishes at discrete times $t_m=2[ m\pi-\arctan(\Omega_n^{'}/\kappa)]/\Omega_n^{'}$ with $m$ integers and $\Omega_n^{'}=\sqrt{4ng^2-\kappa^2}$. However, no stationary entanglement is seen for both coupling regimes. It means that, at sufficiently long times, we are left with an ensemble of non-correlated qubits.

\subsection{Protecting of Entanglement}
Here we consider the action of a series of N nonselective measurements, each performed at time intervals $T=t/N$ in order to check whether the system is still in its initial state. After every measurement, the system is projected back to its initial state and then the temporal evolution starts anew. The survival probability of the initial state after the first observation is $\bra{\psi_0}\rho(T)\ket{\psi_0}=\left| {\cal E}(T)\right| ^2$. The sequence of the N measurements repeatedly brings the system into its initial state with the surviving probability $P_0^{(N)}(t\equiv NT)=\left| {\cal E}(T)\right| ^{2N}$ which can be rewritten after some manipulations as
\begin{equation}
P_0^{(N)}(t)=\exp\left[ -\Gamma_z(T)t\right] ,
\label{eq:proN}
\end{equation}
with an effective decay rate defined as $\Gamma_z(T)=-\log\left[ \left| {\cal E}(T)\right| ^2\right]/T$. It is clear that, for a finite time $t=NT$ and in the limits $T\longrightarrow 0$ and $N\longrightarrow\infty$, one obtains $\Gamma_z(T)\longrightarrow 0$, i.e., the decay completely suppressed. It is clear that, the projective measurements not only affect the probability $P_0(t)$, but also modify the time evolution of the entanglement. More explicitly, according to Eqs. (\ref{eq:suramp}) and (\ref{eq:proN}), the modified concurrence reads as
\begin{equation}
{\cal C}_{\text{pair}}^{(N)}(t)=\dfrac{2\exp\left[ -\Gamma_z(T)t\right]}{n},
\label{eq:modcon}
\end{equation}
whose effective dynamics now depends on $T$. This result can also be directly obtained from the density matrix describing the system has been observed $N$ times, i.e., $\ket{\psi(t)^{(N)}}\bra{^{(N)}\psi(t)}$, by tracing over the reservoir degrees of freedom and over all other qubits.

\begin{figure}[h]
\centering
\subfigure[\label{Fig2a} \ Bad cavity limit, $R=0.1$.]{\includegraphics[width=0.35\textwidth]{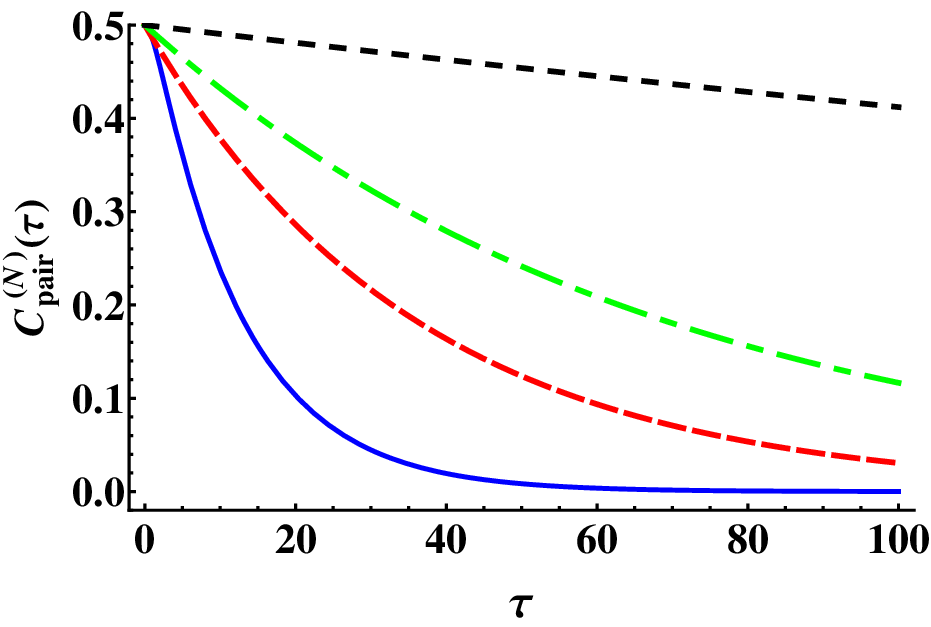}}
\hspace{0.05\textwidth}
\subfigure[\label{Fig2b} \ Good cavity limit, $R=10$.]{\includegraphics[width=0.35\textwidth]{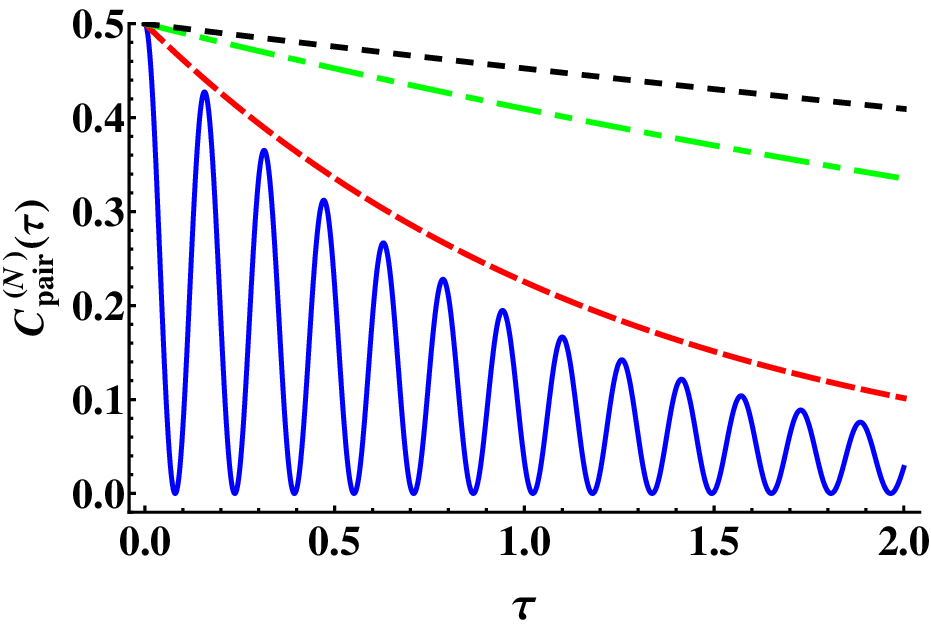}}

\caption{(Color online) Pairwise concurrence as function of $\tau$ for the system size $n=4$, in the absence of measurements (solid blue line) and in the presence of measurements for (a) weak coupling ($R=0.1$) with intervals $\kappa T=5$ (dashed red line), $1$ (dot-dashed green line) and $0.1$ (dotted black line) and (b) strong coupling ($R=10$) with intervals $\kappa T=0.004$ (dashed red line), $0.001$ (dot-dashed green line) and $0.0005$ (dotted black line).} \label{Fig2}
   \end{figure}

Figure \ref{Fig2} illustrates a comparison between the dynamics of pairwise concurrence in the absence and  presence of the nonselective  measurements when the system size is $n=4$ for various intervals $T$. As is clear, the presence of measurements extinguishes the decaying behaviour of the entanglement and also washes out the entanglement sudden death. This decay suppression depends directly on the interval times $T$. Indeed, by decreasing the interval times the system remains in its initial state in longer times. It is worth noting that this quantum Zeno effect desperately depends on the environment features, the resonance condition and the system size. For instance, in the good cavity limit and for large values of interval time $T$, an anti \cite{Ai2013} quantum Zeno \cite{Misra1977} effect may be obtained. This effect may enhance the decay of entanglement. On the other hand, when the system size is increased, smaller interval times is needed to protect entanglement via quantum Zeno effect.

\section{System of  $n$-Qubits Initially in a superposition of one excitation of two arbitrary qubits }\label{sec.twoqubit}
In this section, we  address the case in which the initial state of the qubits are in a superposition of one excitation of two arbitrary qubits. Again, we assume that there is no excitation in the cavity before the occurrence of interaction.  Therefore, the initial state of the whole system+environment can be written as:
\begin{equation}
\ket{\psi(0)}=(c_{01}\ket{1_k}+c_{02}\ket{1_l})\ket{\boldsymbol{0}}_{R},
\label{eq:initialstate}
\end{equation}
in which $\ket{1_i}$ and $\ket{\boldsymbol{0}}_{R}$ have been defined before in the paper. We assume that the initial state is characterized by the separability parameter $s$ as follow:
\begin{equation}\label{eq:c0}
c_{01}=\sqrt{\dfrac{1-s}{2}}, \ \ \ \ \ c_{02}=\sqrt{\dfrac{1+s}{2}}e^{i\varphi},
\end{equation}
with $-1\leq s\leq 1$ and in particular $s=\pm 1$ ($s=0$) corresponds to a  separable (maximum entangled) initial state.
Accordingly, the quantum state of the entire system+environment at any time can be written as

\begin{equation}
\begin{aligned}
\ket{\psi(t)}&=\Big( c_1(t)\ket{1_k}+c_2(t)\ket{1_l}+c_3(t)\ket{E_{\cancel{k} \cancel{l}}}\Big)e^{i\omega_{\text{qb}}t} \ket{\boldsymbol{0}}_{R} \\
 &+\int\! c_{\omega}(t)e^{i\omega t}\ket{1_{\omega}}\ket{G} \, \mathrm{d}\omega,
\end{aligned}
\label{eq:state}
\end{equation}
in which we have defined the normalized state $\ket{E_{\cancel{k}\cancel{l}}}:=\frac{1}{\sqrt{n-2}}\sum_{j\neq k,l}^{n}\ket{1_j}$. Following the procedure presented in obtaining the expression (\ref{eq:survivalamplitude}), one may straightforwardly obtain the following analytical expressions for the time-dependent amplitudes:
\begin{subequations}
\label{eq:usolved}
\begin{eqnarray}
c_{1}(t)&=&\frac{(n-1)c_{01}-c_{02}}{n}+\frac{c_{01}+c_{02}}{n}{\cal E}(t),  \label{eq:usolved1} \\
c_{2}(t)&=&\frac{(n-1)c_{02}-c_{01}}{n}+\frac{c_{01}+c_{02}}{n}{\cal E}(t), \label{eq:usolved2} \\
c_{3}(t)&=&\frac{\sqrt{n-2}}{n}(c_{01}+c_{02})(-1+{\cal E}(t)). \label{eq:usolved3}
\end{eqnarray}
\end{subequations}
From Eq. (\ref{eq:survivalamplitude}), it is clear that letting $t$  to tend to infinity, then ${\cal E}(\infty)\longrightarrow 0$ which leads to the nonzero values of the coefficients $c_i(\infty)$. Therefore, unlike the case with initial Werner state, the environment not only can create entanglement between various pairs of qubits, but also it may make it to persist to be stationary. According to (\ref{eq:usolved}), this stationary state does not depend on the environment features such as the cavity damping rate or coupling constant and only depends on the initial conditions as well as the size of system, $n$. This is due to the fact that we have assumed that the coupling constant be the same for all qubits. It can be shown that, by choosing different coupling constants associated with different qubits, the stationary entanglement depends also on the cavity damping rate and coupling constants.

Using Eq. (\ref{eq:state}), the explicit form of the reduced density operator for the system of qubits at any time can be derived by tracing over environment variables which results in
\begin{equation}
\begin{aligned}
\rho(t)&=\left| c_1(t)\right| ^2\ket{k}\bra{k}+\left| c_2(t)\right| ^2\ket{l}\bra{l}+\left| c_3(t)\right| ^2\ket{E_{\cancel{k}\cancel{l}}}\bra{E_{\cancel{k}\cancel{l}}} \\
&+\left( c_1(t)c_2^{*}(t)\ket{k}\bra{l}+c_1(t)c_3^{*}(t)\ket{k}\bra{E_{\cancel{k}\cancel{l}}} \right. \\
&\left. +c_2(t)c_3^{*}(t)\ket{l}\bra{E_{\cancel{k}\cancel{l}}} + \text{H.c.} \right)\\
&+\left(1-\left| c_1(t)\right| ^2-\left| c_2(t)\right| ^2-\left| c_3(t)\right| ^2 \right) \ket{G}\bra{G}.
\end{aligned}
 \label{eq:densitymatrix}
\end{equation}
In the next two subsections, we shall compare the various pairwise entanglements resulting from initially two qubits ($k$th and $l$th) which are in a superposition of maximally entangled state (i.e., $s=0$) and when only one qubit (namely $k$th qubit) is initially in the excited state (i.e., $s=-1$).

\subsection{Maximum Entangled Initial State}\label{sec:maxini}
In this subsection, we assume that the system of qubits be initially in a maximum entangled state of two qubits (namely $k$th and $l$th). This can easily be done by setting $s=0$ and {$\varphi=0$} in (\ref{eq:c0}).
To analyze the pairwise entanglement between qubits $k$ and $l$ (initially in a superposition of excited states), we compute partial trace of (\ref{eq:densitymatrix}) over all other qubits and obtain
\begin{equation}
\begin{aligned}
\rho_{k,l}(\tau)&=\left| c_1(\tau)\right| ^2\ket{10}\bra{10}+\left| c_2(\tau)\right| ^2\ket{01}\bra{01} \\
&+ c_1(\tau)c_2^{*}(\tau)\ket{10}\bra{01}+c_1^{*}(\tau)c_2(\tau)\ket{01}\bra{10} \\
&+\left(1-\left| c_1(\tau)\right| ^2-\left| c_2(\tau)\right| ^2 \right) \ket{00}\bra{00},
\end{aligned}
 \label{eq:rkl}
\end{equation}
which leads to the following concurrence:
\begin{equation}
{\cal C}_{k,l}(\tau)=2\left| c_1(\tau)\right| \left| c_2(\tau)\right| .
\label{eq:ckl}
\end{equation}
\begin{figure}[h]
\centering
\subfigure[\label{Fig3a} \ Bad cavity limit, $R=0.1$.]{\includegraphics[width=0.35\textwidth]{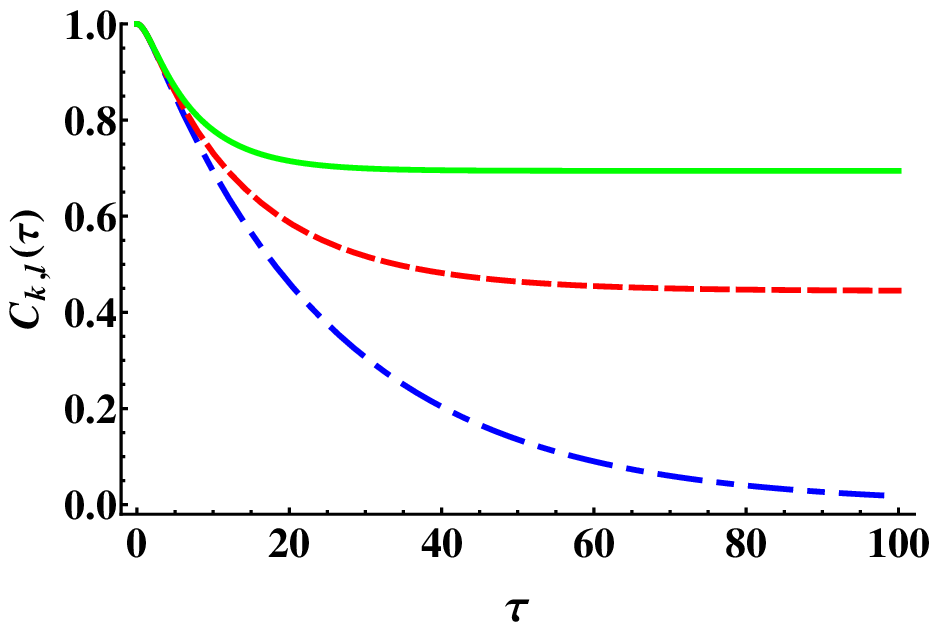}}
\hspace{0.05\textwidth}
\subfigure[\label{Fig3b} \ Good cavity limit, $R=10$.]{\includegraphics[width=0.35\textwidth]{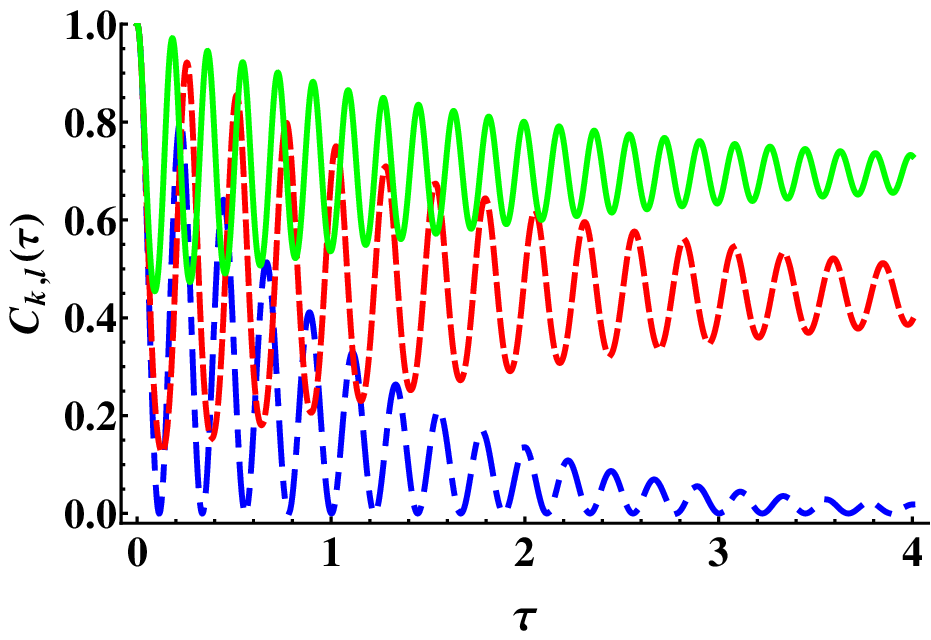}}

\caption{(Color online) Pairwise concurrence ${\cal C}_{k,l}$ as function of $\tau$ for $s=\varphi=0$ in the bad cavity limit, i.e. $R=0.1$ (top plots) and good cavity limit, $R=10$ (bottom plots) with $n=2$ (dot-dashed blue lines), $n=6$ (dashed red lines),  and $n=12$ (solid green lines).} \label{Fig3}
   \end{figure}
Figure \ref{Fig3} illustrates the time evolution of the concurrence ${\cal C}_{k,l}(\tau)$ as a function of the scaled time $\tau$ for weak and strong coupling regimes for a maximally entangled initial state (i.e., $s=0$ and $\varphi=0$). In the weak coupling regime, concurrence falls down from its maximum initial value and monotonically decreases until it reaches its stationary value. In the strong coupling regime, an oscillatory behaviour along with decaying of entanglement is clearly seen such that for $n=2$ the entanglement sudden death is occurred. As it is mentioned before, both strong and weak coupling regimes lead to the same stationary state. The surprising aspect here is that for $n=2$ the entanglement between two qubits vanishes under the contamination of the environment, but adding more number of qubits maintains the entanglement stored between these two qubits. In general, when the system size $n$ becomes larger, the stationary entanglement increases and the concurrence achieves sooner its stationary value.

Letting $t$ to tend to infinity in Eq. (\ref{eq:ckl}), we found the behaviour of stationary entanglement versus system size $n$ as
\begin{equation}\label{eq:stackl}
{\cal C}_{k,l}(\infty)=\dfrac{(n-2)^2}{n^2}.
\end{equation}

With the help of Eq. (\ref{eq:densitymatrix}), we compute the reduced density operator between $k$th qubit and another generic  qubit $j$ (initially in the ground state) as:
\begin{equation}
\begin{aligned}
\rho_{k,j}(\tau)&=\left| c_1(\tau)\right| ^2\ket{10}\bra{10}+\frac{\left| c_3(\tau)\right| ^2}{n-2}\ket{01}\bra{01} \\
&+ \frac{1}{\sqrt{n-2}}\left( c_1(\tau)c_3^{*}(\tau)\ket{10}\bra{01}+c_1^{*}(t)c_3(\tau) \ket{01}\bra{10}\right)  \\
&+\left(1-\left| c_1(\tau)\right| ^2-\frac{\left| c_3(\tau)\right| ^2}{n-2} \right) \ket{00}\bra{00}.
\end{aligned}
 \label{eq:rkj}
\end{equation}
At last, the relevant concurrence using (\ref{eq:rkj}) results in
\begin{equation}
C_{k,j}(\tau)=\frac{2}{\sqrt{n-2}}\left| c_1(\tau)\right| \left| c_3(\tau)\right|,
\label{eq:ckj}
\end{equation}
which is valid for $n>2$. According to (\ref{eq:ckj}) it can be shown that the pairwise entanglement starts from zero and increases up to its stationary entanglement. The stationary entanglement can be determined from Eq. (\ref{eq:ckj}) by letting $t$ going to infinity as follow
\begin{equation}\label{eq:stackj}
{\cal C}_{k,j}(\infty)=\dfrac{2(n-2)}{n^2}.
\end{equation}
It is obvious that, the maximum  stationary entanglement ${\cal C}_{k,j}^{(max)}=0.25$ is achieved for system size $n=4$.

\begin{figure}[H]
\centering
\subfigure[\label{Fig4a} \ Bad cavity limit, $R=0.1$.]{\includegraphics[width=0.35\textwidth]{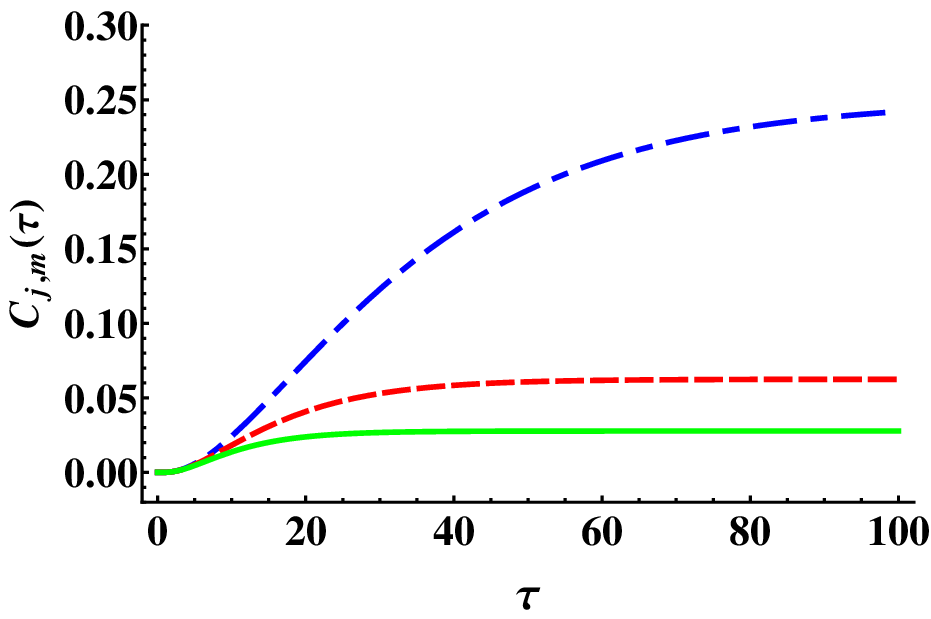}}
\hspace{0.05\textwidth}
\subfigure[\label{Fig4b} \ Good cavity limit, $R=10$.]{\includegraphics[width=0.35\textwidth]{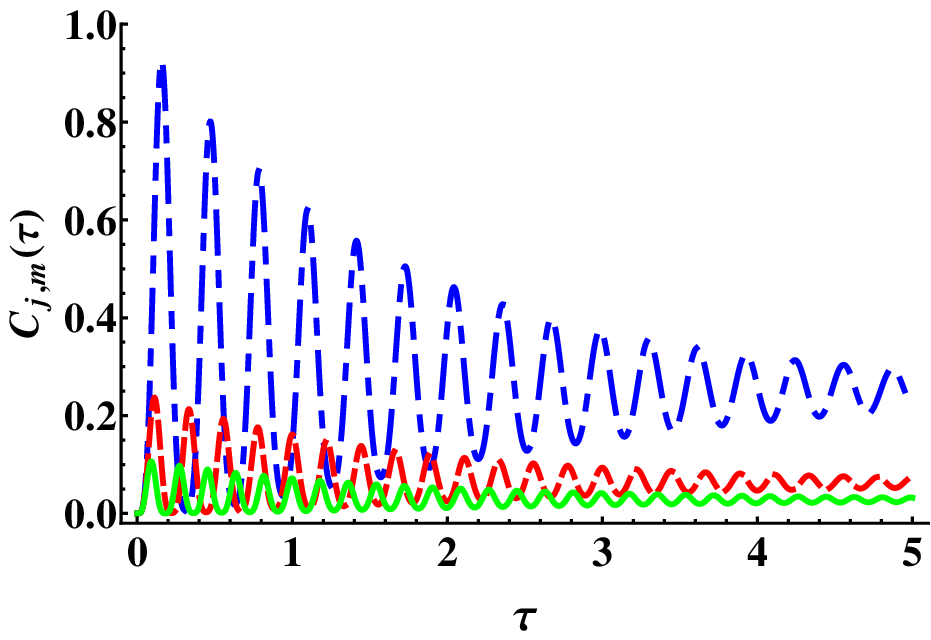}}
\caption{(Color online) Pairwise concurrence ${\cal C}_{j,m}$ as function of $\tau$ for $s=\varphi=0$ in the bad cavity limit, i.e. $R=0.1$ (top plots) and good cavity limit, $R=10$ (bottom plots) with $n=4$ (dot-dashed blue lines), $n=8$ (dashed red lines), $n=12$ (solid green lines).} \label{Fig4}
   \end{figure}
The other possible case which we study is the entanglement between two generic qubits $j$ and $m$ initially in the ground state  ($j,m\neq k,l$). The corresponding reduced density operators reads as:
\begin{equation}
\begin{aligned}
\rho_{j,m}(\tau)&=\frac{\left| c_3(\tau)\right| ^2}{n-2}\left(\ket{10}+\ket{01}\right) \left( \bra{10}+\bra{01}\right)   \\
&+\left(1-2\frac{\left| c_3(\tau)\right| ^2}{n-2} \right) \ket{00}\bra{00}.
\end{aligned}
 \label{eq:rjm}
\end{equation}
The corresponding concurrence can be easily obtained as
\begin{equation}
C_{j,m}(\tau)=\frac{2}{n-2}\left| c_3(\tau)\right|^2 ,
\label{eq:cjm}
\end{equation}
which is valid for $n>2$. A glance at Fig. \ref{Fig4} provided reveals the dynamical behaviour of the $C_{j,m}(\tau)$ in the bad and good cavity limits for $s=0$. It is evident that the entanglement sudden death phenomenon has occurred in the good cavity limit. It is apparent from the information supplied that, in the latter regime, the amount of revived entanglement has become considerably comparable to 1 at short times and for small system sizes. It is also interesting to notice that, both coupling regimes lead to the same stationary entanglement which vanishes for large system sizes. In fact, by letting $t$ go to infinity in Eq. \eqref{eq:cjm} and computing the stationary concurrence as $C_{j,m}(\infty)=\frac{4}{n^2}$, one can easily observe that, for large system sizes, the latter concurrence is by far more negligible compared to the other stationary concurrences.

Altogether, by comparing various stationary entanglements which have been appeared, it can be concluded that when the system of qubits  initially is in the maximum entangled state of two qubits, we have the graph depicted in Fig. {\ref{Fig5} as the steady state. The ticker line in Fig. \ref{Fig5} implies the fact that, at the steady state, the correlation between the initially excited qubits is stronger than the correlation between any other two qubits.

\begin{figure}[H]
\centering
\includegraphics[width=0.3\textwidth]{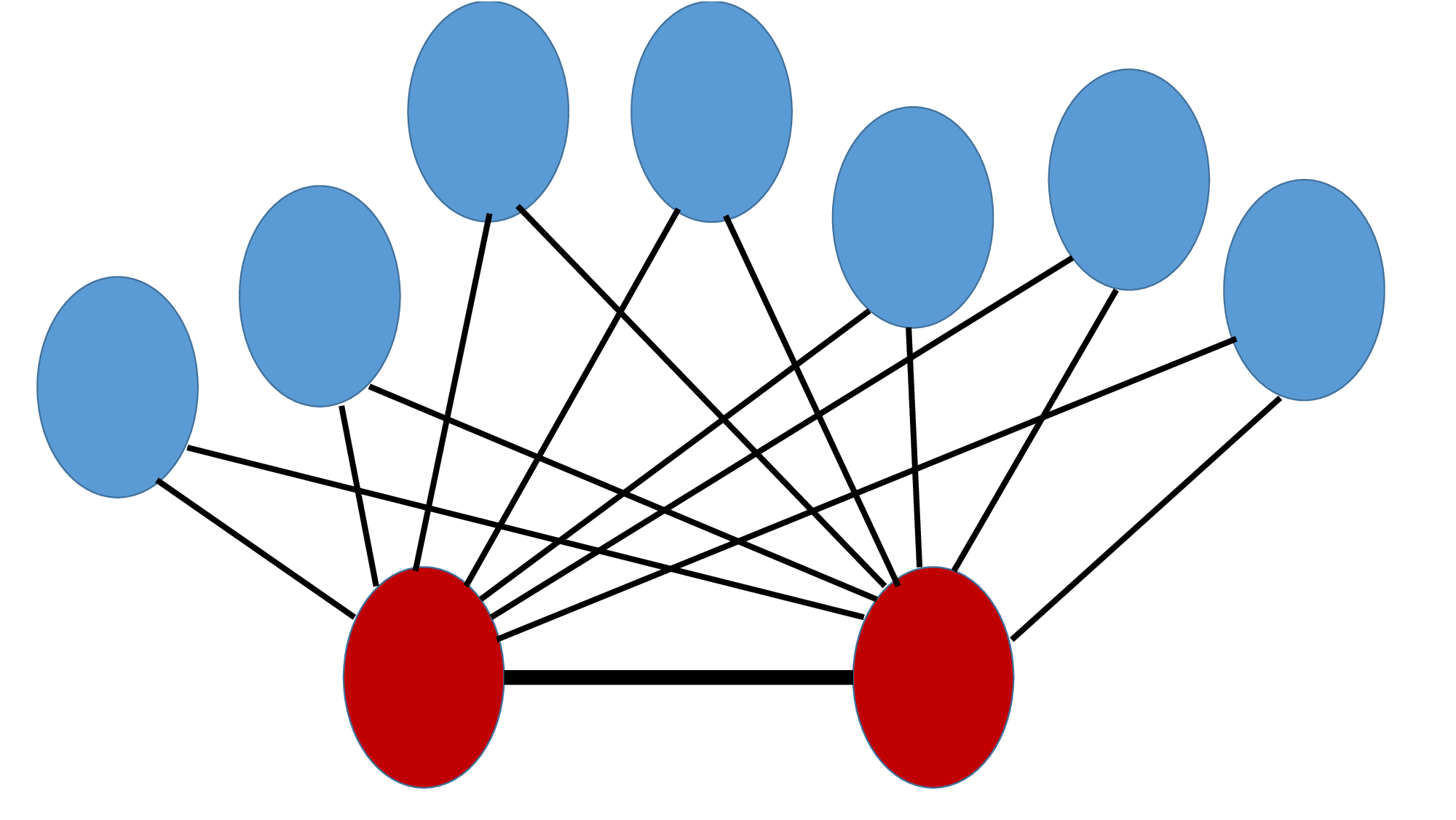}
\caption{(Color online) Pictorial representation of the leading stationary correlations from an initially Werner state. Red (blue) circles represent qubits initially in the superposition of excited states (ground states). Solid lines represent the quantum correlations between qubits at steady state.} \label{Fig5}
\end{figure}

\subsection{One Initial Excitation}{\label{sec:oneini}
In this subsection, we assume that only $k$th qubit is initially in the excited state (i.e., $s=-1$). In order to analyse the pairwise entanglement between qubit $k$ and another generic qubit $j$, it is enough to set $s=-1$ in  Eqs. (\ref{eq:ckl}) or (\ref{eq:ckj}). The dynamical behaviour of ${\cal C}_{\text{k,j}}(\tau)$ is shown in Fig. \ref{Fig6} in both strong and weak coupling regimes for some values of system size $n$. It is easy to show that at the steady state, the pairwise concurrence takes the form
\begin{equation}\label{eq:stackjs-1}
{\cal C}_{k,j}(\infty)=\dfrac{2(n-1)}{n^2}.
\end{equation}
On the other hand, the entanglement between two other generic qubits, initially in the ground state, has similar behaviour to Fig. \ref{Fig4}. In particular, its corresponding stationary concurrence takes the form $C_{j,m}(\infty)=\frac{2}{n^2}$ which vanishes for large system sizes. Therefore, we have a star graph as the steady state (see Fig. \ref{Fig7}).
\begin{figure}[h]
\centering
\subfigure[\label{Fig6a} \ Bad cavity limit, $R=0.1$ with $\varphi=0$.]{\includegraphics[width=0.35\textwidth]{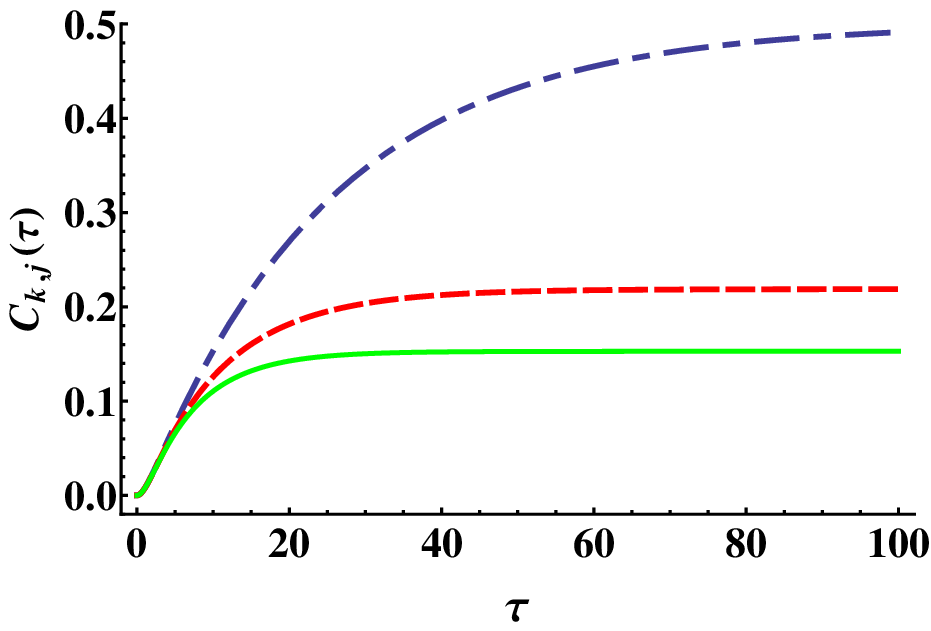}}
\hspace{0.05\textwidth}
\subfigure[\label{Fig6b} \ Good cavity limit, $R=10$ with $\varphi=0$.]{\includegraphics[width=0.35\textwidth]{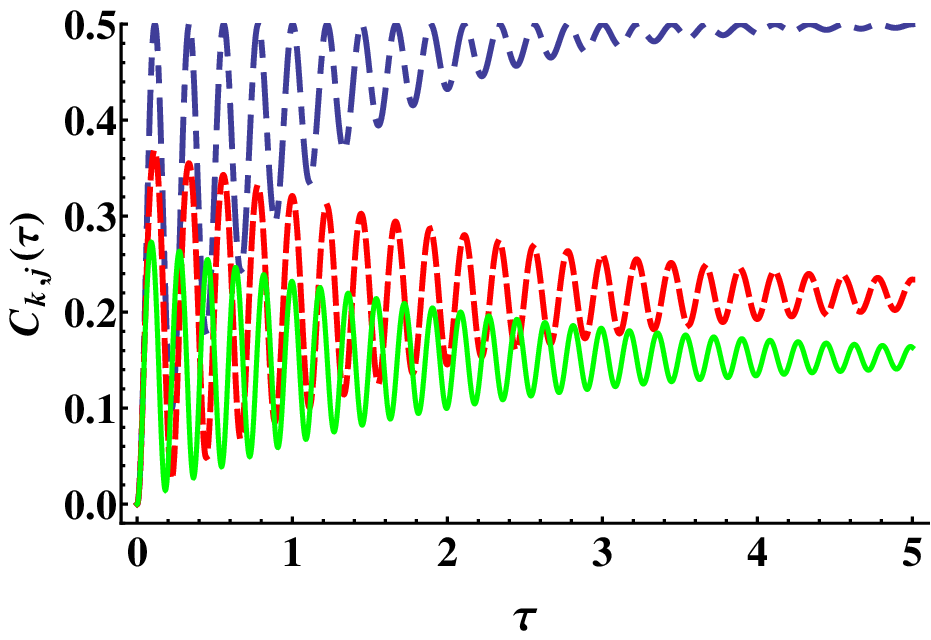}}
\caption{(Color online) Pairwise concurrence ${\cal C}_{k,j}$ as function of $\tau$ for $s=-1$ in the bad cavity limit, i.e. $R=0.1$ (top plots) and good cavity limit, $R=10$ (bottom plots) with $n=4$ (dot-dashed blue lines), $n=8$ (dashed red lines), $n=12$ (solid green lines).} \label{Fig6}
   \end{figure}

\begin{figure}[H]
\centering
\includegraphics[width=0.3\textwidth]{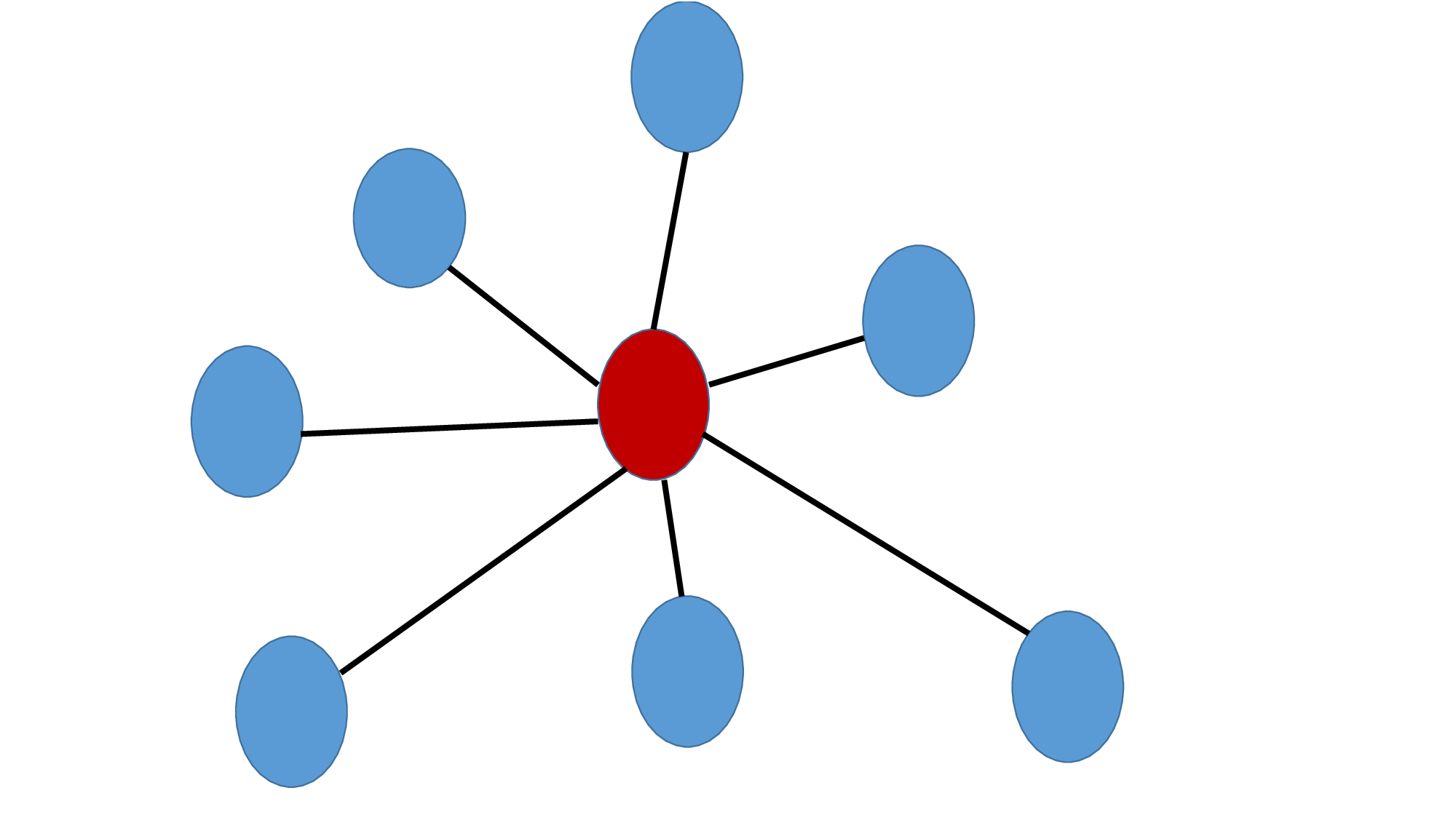}
\caption{(Color online) Pictorial representation of the leading stationary concurrence when initially only one qubit is in the excited state. Solid lines represent the quantum correlations between qubits at steady state.} \label{Fig7}
\end{figure}

\section{Concluding Remarks}\label{sec.conclusion}
To sum up, we have studied an exactly solvable model describing an arbitrary number of qubits dissipating into a common environment where both Markovian and non-Markovian effects corresponding to the bad and good cavity limits, respectively, are present. In the weak coupling regime the pairwise entanglement decays (and sometimes increments) exponentially and goes up to its stationary value only asymptotically. On the other hand, in the strong coupling regime, the memory effect of the environment allows entanglement to oscillate before a stationary value is reached.

We  found that, for initially a Werner state, the pairwise entanglement has a decaying behaviour with no stationary value for both strong and weak coupling regimes. However, we showed that a series of non-selective measurements can preserve the entanglement initially stored in the system of qubits which is known as quantum Zeno effect. Moreover, it should be noted that, this effect depends on the time intervals, the environment features as well as the presence or  absence of detuning. An anti-Zeno quantum effect which enhances the decay of entanglement can be occurred for some situations. This investigation is left for future work.

We also studied the possibility of achieving stationary pairwise entanglement states by limiting the number of qubits initially in the superposition of only one excitation of qubits. More specially, we addressed the case in which the system  is initially in the superposition of one excitation of two arbitrary qubits. In this case, the stationary entanglement is independent of environment properties and depends on the system size $n$ and the initial conditions which is characterized by the separability parameter $s$. This is because of the assumption that the coupling constant of all qubits are the same. Although the interaction of the system-environment leads to vanishing of the entanglement for system size $n=2$ with Bell state as its initial state, as an interesting result, increasing the number of qubits satisfactorily preserves the initial entanglement (see Fig. \ref{Fig3}).

The stationary pairwise entanglement ${\cal C}_{k,l}(\infty)$ (here $k$th and $l$th qubits  are initially in the superposition of one excitation, see Eq. (\ref{eq:initialstate})) monotonically increases with the system size $n$. It is even possible to achieve the maximum pairwise entanglement in enough large system sizes.

The entanglement can also be created and persists at a steady state for pairs of initially excited and non-excited qubits (see Fig. \ref{Fig6}). It is also possible to create entanglement between pairs of qubits initially in the ground state (see Fig. \ref{Fig4}). However, in such a case the amount of entanglement is negligibly smaller than the previous cases and also is nearly independent of separability parameter.

It is worth noticing that, when only one qubit is initially in the excited state (i.e., $s=1$ or $-1$), we have a star graph as the steady state for large systems in both weak and strong coupling regimes. This is quite in consistent with previous works (see for example, \cite{Memarzadeh2013}). On the other hand, when two qubits  are initially in a maximum entangled state, we are left with a bipartite graph with strong correlation between the two qubits (which  are initially in a maximum entangled state). This differs from the case where initially two qubits are in the excited states simultaneously. In this case, at the steady state we have a bipartite graph but without any correlation between the two qubits which  are initially excited \cite{Memarzadeh2013}. Altogether, this subject can be of interest from the perspective of quantum complex networks \cite{Perseguers2010}.

The observed aspects  could be verified and confirmed in experiments which focused on trapped ions coupled to the dissipative bath of vacuum modes of the radiation field via optical pumping \cite{Barreiro2011}. Also, the system of superconducting Josephson circuits as qubits and a transmission line as cavity \cite{Majer2007} could be a suitable candidate as an experimental implementation to explore the contents of the this work. The present work can also be relevant for driving cavity QED experiments with noninteracting qubits inside a cavity \cite{Specht2011}. Besides, the continuous miniaturization of physical devices compels us to consider dissipative models with a common environment, regardless of the presence or absence of direct subsystem interactions \cite{Pulido2008}. Therefore, we expect the presented analytical results would be the first step towards that goal.


\begin{thebibliography}{10}
\bibitem{Horodecki2009QuantumEntanglement} R. Horodecki, P. Horodecki, M. Horodecki, and K. Horodecki, Rev. Mod. Phys. \textbf{81}, 865 (2009).

\bibitem{Ekert1991Cryptography} A. K. Ekert, Phys. Rev. Lett. \textbf{67}, 661 (1991).

\bibitem{Braunstein1995} S. L. Braunstein and A. Mann, Phys. Rev. A. \textbf{51}, R1727 (1995).

\bibitem{Mattle1996} K. Mattle, H. Weinfurter, P. G. Kwiat, and A. Zeilinger, Phys. Rev. Lett. \textbf{76}, 4656 (1996).

\bibitem{Abdi2012} M. Abdi, S. Pirandola, P. Tombesi, and D. Vitali, Phys. Rev. Lett. \textbf{109}, 143601 (2012).

\bibitem{Ou2012} Z. Y. Ou, Phys. Rev. A \textbf{85}, 023815 (2012).

\bibitem{Muarao1999} M. Murao, D. Jonathan, M. B. Plenio, and V. Vedral, Phys. Rev. A \textbf{59}, 156 (1999).

\bibitem{Plenio2002} M. B. Plenio and S. F. Huelga, Phys. Rev. Lett. \textbf{88}, 197901 (2002).

\bibitem{Mancini2015} S. Mancini and J. Wang, Eur. Phys. J. D \textbf{32}, 257 (2005).

\bibitem{Angelakis2009} D. G. Angelakis, S. Bose, and S. Mancini, Europhys. Lett. \textbf{85}, 20007 (2009).

\bibitem{Krauter2011} H. Krauter, C. A. Muschik, K. Jensen, W. Wasilewski, J. M. Petersen, J. I. Cirac, and E. S. Polzik, Phys. Rev. Lett. \textbf{107}, 080503 (2011).

\bibitem{Nourmandipour2015} A. Nourmandipour and M. K. Tavassoly, J. Phys. B: At., Mol. Opt. Phys. \textbf{48}, 165502 (2015).

\bibitem{Memarzadeh2013} L. Memarzadeh and S. Mancini, Phys. Rev. A \textbf{87}, 032303 (2013).

\bibitem{Maniscalco2008} S. Maniscalco, F. Francica, R. L. Zaffino, N. Lo Gullo, and F. Plastina,  Phys. Rev. Lett.  \textbf{100}, 090503 (2008).

\bibitem{Memarzadeh2011} L. Memarzadeh and S. Mancini, Phys. Rev. A \textbf{83}, 042329 (2011).

\bibitem{Rafiee2012} M. Rafiee, C. Lupo, H. Mokhtari, and S. Mancini, Phys. Rev. A \textbf{85}, 042320 (2012).

\bibitem{Vasile2009} R. Vasile, S. Olivares, M. G. A. Paris, and S. Maniscalco, Phys. Rev. A \textbf{80}, 062324 (2009).

\bibitem{Genkin2011} M. Genkin and A. Eisfeld, J. Phys. B: At., Mol. Opt. Phys. \textbf{44}, 035502 (2011).

\bibitem{Gao2010} G.- xiang Li, L.- hui Sun, and Z. Ficek, J. Phys. B: At., Mol. Opt. Phys. \textbf{43}, 135501 (2010).

\bibitem{Paz2008} J. P. Paz and A. J. Roncaglia, Phys. Rev. Lett. \textbf{100}, 220401 (2008).

\bibitem{Horhammer2008} C. H\"{o}rhammer and H. B\"{u}ttner, Phys. Rev. A \textbf{77}, 042305 (2008).

\bibitem{Kim2012} Y.-S. Kim, J.-C. Lee, O. Kwon, and Y.-H. Kim, Nat. Phys. \textbf{8}, 117 (2012).

\bibitem{Ghanbari2014} R. Ghanbari and M. Rafiee, Eur. Phys. J. D \textbf{68}, 215 (2014).

\bibitem{Kofman2000} A. Kofman and G. Kurizki, Nature (London) \textbf{405}, 546 (2000).

\bibitem{Facchi2002} P. Facchi and S. Pascazio, Phys. Rev. Lett. \textbf{89}, 080401 (2002).

\bibitem{Dutra2005} S. M. Dutra, \emph{Cavity Quantum Electrodynamics: The Strange Theory of Light in a Box} (Wiley, New York, 2005).

\bibitem{Fano1961} U. Fano, Phys. Rev. \textbf{124}, 1866 (1961).

\bibitem{Barnett2002} S. M. Barnett and P. M. Radmore, \emph{Methods in Theoretical Quantum Optics}, Vol. 15 (Oxford University Press, New York, 2002).

\bibitem{Werner1989} R. F. Werner, Phys. Rev. A \textbf{40}, 4277 (1989).

\bibitem{Breuer2002} H.-P. Breuer and F. Petruccione, \emph{The Theory of Open Quantum Systems} (Oxford University Press, New York, 2002).

\bibitem{Bellomo2007} B. Bellomo, R. Lo Franco, and G. Compagno, Phys. Rev. Lett. \textbf{99}, 160502 (2007).

\bibitem{Tavis1968} M. Tavis and F. W. Cummings, Phys. Rev. \textbf{170}, 379 (1968).

\bibitem{Wootters1998} W. K. Wootters, Phys. Rev. Lett. \textbf{80}, 2245 (1998).

\bibitem{Ai2013} Q. Ai, D. Xu, S. Yi, A. Kofman, C. P. Sun, and F. Nori, Sci. Rep. \textbf{3}, 1752 (2013).

\bibitem{Misra1977} B. Misra and E. C. G. Sudarshan, J. Math. Phys. \textbf{18}, 756 (1977).

\bibitem{Perseguers2010} S. Perseguers, M. Lewenstein, A. Ac\'{i}n, and J. I. Cirac, Nat. Phys. \textbf{6}, 539 (2010).

\bibitem{Barreiro2011} J. T. Barreiro, M. Muller, Ph. Schindler, D. Nigg, Th. Monz, M. Chwalla, M. Hennrich, Ch. F. Roos, P. Zoller, and R. Blatt, Nature (London) \textbf{470}, 486 (2011).

\bibitem{Majer2007} J. Majer et al., Nature (London) \textbf{449}, 443 (2007).

\bibitem{Specht2011} H. P. Specht, Ch. Nolleke, A. Reiserer, M. Uphoff, E. Figueroa, S. Ritter, and G. Rempe, Nature (London) \textbf{473}, 190 (2011).

\bibitem{Pulido2008} L. D. Contreras-Pulido and R. Aguado, Phys. Rev. B \textbf{77}, 155420 (2008).

\end{thebibliography}
\end{document}